\documentclass[%
 aip,
 amsmath,amssymb,
 reprint,%
]{revtex4-2}

\usepackage{graphicx}
\usepackage{dcolumn}
\usepackage{bm}
\usepackage[utf8]{inputenc}
\usepackage[T1]{fontenc}
\usepackage{mathptmx}
\usepackage{etoolbox}
\usepackage{calc}
\usepackage{comment}
\usepackage{graphicx}
\usepackage{caption}
\usepackage{subcaption}
\usepackage[shortlabels]{enumitem}
\usepackage{amsmath}
\usepackage{amsfonts}
\usepackage{amssymb}
\usepackage{float}
\usepackage{xcolor}
\usepackage{spalign}
\usepackage{natbib}
\usepackage{systeme}
\usepackage{units}

\DeclareMathOperator\im{Im}

\makeatletter
\def\@email#1#2{%
 \endgroup
 \patchcmd{\titleblock@produce}
  {\frontmatter@RRAPformat}
  {\frontmatter@RRAPformat{\produce@RRAP{*#1\href{mailto:#2}{#2}}}\frontmatter@RRAPformat}
  {}{}
}%
\makeatother
\begin{document}

\preprint{AIP/123-QED}

\title[]{Terahertz control of photoluminescence emission in few-layer InSe}

\author{T. Venanzi} %
 \email{tommaso.venanzi@uniroma1.it}
 \affiliation{Helmholtz-Zentrum Dresden-Rossendorf, 01314 Dresden, Germany} 
 \affiliation{Dipartimento di Fisica, Università di Roma "Sapienza", 00185 Rome, Italy}
\author{M. Selig}
\affiliation{Institut f\"ur Theoretische Physik, Technische Universit\"at Berlin, 10623 Berlin, Germany}
\author{A. Pashkin}
 \affiliation{Helmholtz-Zentrum Dresden-Rossendorf, 01314 Dresden, Germany}
\author{S. Winnerl}
 \affiliation{Helmholtz-Zentrum Dresden-Rossendorf, 01314 Dresden, Germany}
 \author{M. Katzer}
\affiliation{Institut f\"ur Theoretische Physik, Technische Universit\"at Berlin, 10623 Berlin, Germany}
\author{H. Arora}
 \affiliation{Helmholtz-Zentrum Dresden-Rossendorf, 01314 Dresden, Germany}
\author{A. Erbe}
 \affiliation{Helmholtz-Zentrum Dresden-Rossendorf, 01314 Dresden, Germany}
\author{A. Patan{\`{e}}}
\affiliation{School of Physics and Astronomy, University of Nottingham, Nottingham NG7 2RD, UK}
\author{Z. R. Kudrynskyi}
\affiliation{School of Physics and Astronomy, University of Nottingham, Nottingham NG7 2RD, UK} 
\author{Z. D. Kovalyuk}
\affiliation{Frantsevich Institute for Problems of Materials Science, The National Academy of Sciences of Ukraine, Chernivtsi Branch, 58001 Chernivtsi, Ukraine} 
\author{L. Baldassarre}
 \affiliation{Dipartimento di Fisica, Università di Roma "Sapienza", 00185 Rome, Italy}
\author{A. Knorr}
\affiliation{Institut f\"ur Theoretische Physik, Technische Universit\"at Berlin, 10623 Berlin, Germany}
\author{ M. Helm}
 \affiliation{Helmholtz-Zentrum Dresden-Rossendorf, 01314 Dresden, Germany}
 \affiliation{Technische Universit\"at Dresden, 01062 Dresden, Germany}
\author{H. Schneider}
 \affiliation{Helmholtz-Zentrum Dresden-Rossendorf, 01314 Dresden, Germany}

\date{\today}

\begin{abstract}
A promising route for the development of opto-elelctronic technology is to use terahertz radiation to modulate the optical properties of semiconductors. Here we demonstrate the dynamical control of photoluminescence (PL) emission in few-layer InSe using picosecond terahertz pulses. We observe a strong PL quenching (up to $50\%$) after the arrival of the terahertz pulse followed by a reversible recovery of the emission on the time scale of $50\,$ps at $T=10\,$K. Microscopic calculations reveal that the origin of the photoluminescence quenching is the terahertz absorption by photo-excited carriers: this leads to a heating of the carriers and a broadening of their distribution, which reduces the probability of bimolecular electron-hole recombination and, therefore, the luminescence. By numerically evaluating the Boltzmann equation, we are able to clarify the individual roles of optical and acoustic phonons in the subsequent cooling process. The same PL quenching mechanism is expected in other van der Waals semiconductors and the effect will be particularly strong for materials with low carrier masses and long carrier relaxation time, which is the case for InSe. This work gives a solid background for the development of opto-electronic applications based on InSe, such as THz detectors and optical modulators.
\end{abstract}

\maketitle

The use of terahertz radiation to modulate the optical properties of van der Waals (vdW) semiconductors is a promising route for the development of opto-electronic technology \cite{Langer2018}. Indeed this class of materials offers unique engineering possibilities by controlling the material thickness down to monolayers and by making heterostructures without lattice matching constrains and with the additional twist-angle degree of freedom \cite{Geim2013,Ubrig2020,Liu2019}. Among vdW materials, InSe is a III-VI semiconductor that has shown numerous promising properties for opto-electronic applications: it features a direct band gap, low electron mass that leads to high electron mobility \cite{Bandurin2017,Sucharitakul2015} and large band-gap tunability \cite{Mudd2013,Li2018,Song2018}, and very low Young's modulus that leads to high plasticity \cite{Zhao2019,Wei2020}. Therefore, InSe can be used as active layer for various applications like photodetectors \cite{Tamalampudi2014,Lei2014}, field-effect transistors \cite{Feng2014,Sucharitakul2015}, and flexible electronics \cite{Tamalampudi2014,Dai2019,Wu2020}. 

It was recently shown that InSe is an appealing material for applications in the THz range such as detectors and emitters. In fact, intersubband electronic transitions in few-layer InSe were predicted and observed experimentally by means of electronic resonant tunneling \cite{Magorrian2018,Zultak2020,Kudrynskyi2020}. However, the use of InSe in the infrared and terahertz range of the electromagnetic spectrum has not been intensively investigated so far. On one side, it is of fundamental interest to investigate the THz/infrared response of InSe \cite{lauth2016photogeneration,lu2020terahertz}, which in turn could constitute a novel characterization technique for InSe quality testing to be employed in the more developed visible range technology. On the other side, THz radiation can be employed to control the optical response of the material and the light-matter interaction \cite{kampfrath2013resonant,venanzi2021terahertz}. 

Here we demonstrate the dynamical control of photoluminescence emission in InSe using picosecond terahertz pulses obtained with the free-electron laser (FEL) FELBE. To this end, we perform a two-color time-resolved photoluminescence (PL) experiment. After exciting an electron-hole population with picosecond visible (VIS) pulses, a THz pulse arrives with a delay on the sample causing a transient quenching of the PL emission. For all the investigated temperatures, we observe a full recovery of the PL intensity, e.g. on the time scale of $50\,$ps at $T=10\,$K. Our microscopic analysis reveals that the transient quenching is due to the heating of the carriers after THz absorption: this leads to a broadening of the electron and hole distributions in the momentum space that reduces the rate of spontaneous emission. By monitoring the subsequent recovery of the PL, we are able to extract the effective cooling times. We accompany our experiment with a microscopic calculation of carrier cooling obtaining good agreement. The present study is important for the basic physics knowledge of the electron-phonon dynamics in InSe and for the development of THz and opto-electronic applications as ultrafast THz detectors, optical modulators, and ultrafast lasing switchers.

The InSe samples are fabricated via mechanical exfoliation and are transferred onto a diamond substrate. The thicknesses of the flakes are measured with atomic force microscopy and the data are reported in Supplementary Material (SM). The InSe flakes under investigation have a thickness between $8$ to $40\,$nm. If it is not stated differently, we show results from a $16.5\,$nm thick sample since the experimental results do not show any significant thickness dependence. This corresponds to about $20$ atomic layers \cite{Mudd2016}. In this work, we focus on flakes thicker than $8\;$nm since they feature strong PL emission due to the direct band gap \cite{Mudd2016}. For this thickness, the band-gap energy of InSe is of about $1.31\;$eV.

Time-resolved photoluminescence spectra are recorded with a streak camera coupled to a spectrometer. With this system $5\,$ps time resolution and $1\,$meV spectral resolution are obtained. A mode-locked Ti:sapphire (TiSa) oscillator with pulse length of $3\,$ps is used as excitation source at $\lambda=770\,$nm. The spot diameter is around $3\,\mu$m. The THz radiation is obtained with an infrared FEL (FELBE). The THz frequency is tuned from  $5.5$ to $26\,$THz (from $22$ to $110\,$meV). The THz pulse length is of about $5\,$ps. The TiSa laser is reduced from $78\,$MHz to $13\,$MHz by pulse-picking with a Pockels cell to match the FEL repetition rate.

\begin{figure*}[htp]
\centering
\includegraphics[scale=0.6]{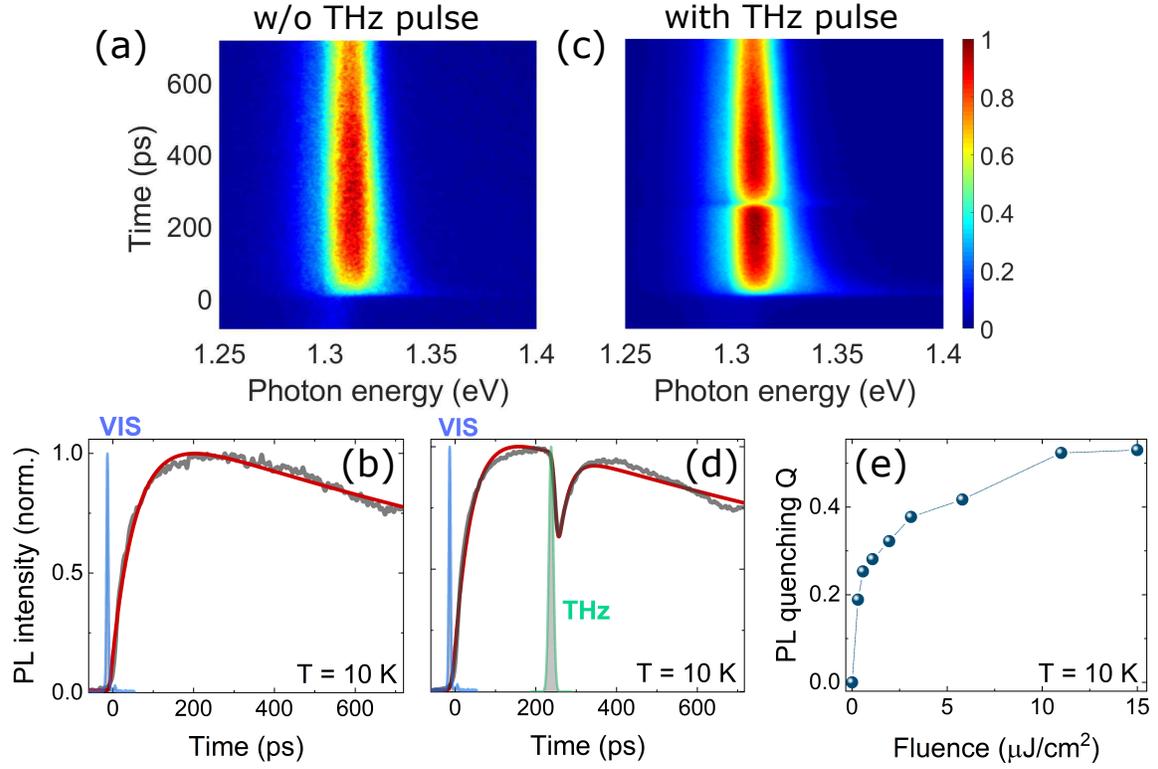}

\caption{(a) Time-resolved PL spectrum of a thin-flake of InSe at $T=10\,$K. (b) Time dependence of the PL signal integrated over the emission photon energy (dark grey curve). The red curve is a fit of the experimental data. (c) Time-resolved PL spectrum with additional THz pumping. The THz pulse arrives $250\,$ps after the optical pulse and induces a transient PL quenching. The FEL fluence is $\phi_{FEL} =5\,\mu$J/cm$^2$ at $12\,$THz. (d) Time dependence of the integrated PL signal with THz radiation. (e) THz fluence dependence of the quenching parameter $Q$. }
\label{fig:trplFEL}
\end{figure*}

Figure \ref{fig:trplFEL}(a) shows the time-resolved photoluminescence spectrum without THz pumping. The visible radiation excites electrons well-above the band gap. After the initial thermalization and relaxation of the photo-excited carriers, a long-living PL emission is observed at $1.31\,$eV. The PL emission is assigned to a combination of defect-assisted radiative recombination and band-to-band recombination\cite{Mudd2014,Venanzi2020}. Figure \ref{fig:trplFEL}(b) shows the PL decay integrated over the photon energy. The PL decay time of the PL emission is longer than the time range of our setup, i.e. $\tau_D>1\,$ns.

Figure \ref{fig:trplFEL}(c) shows the time-resolved PL spectrum with the additional THz pulse. The THz pulse arrives around $250\;$ps after the visible pulse and induces a PL quenching. Shortly after the quenching, the PL emission is recovered. 

The time dependence of the spectrally integrated PL emission is shown in figure \ref{fig:trplFEL}(d). The PL emission is quenched by about $35\%$ by the THz radiation at THz fluence $\phi_{FEL} =5\,\mu$J/cm$^2$ and reaches values higher than $50\%$ at higher fluences (see figure \ref{fig:trplFEL}(e)). The THz fluence dependence of the PL quenching parameter $Q$ shows a saturation behavior at high fluences (Fig. \ref{fig:trplFEL}(e)). The quenching parameter $Q$ is the fraction of PL quenching during the THz pulse arrival with respect to the PL just before the THz pulse. It ranges between 0 (no quenching) and 1 (full quenching).

In order to extract the PL decay constants, a system of rate equations is used containing the decay constants ($\tau_{rise}$, $\tau_{PL}$, and $\tau_{cooling}$) and the quenching parameter $Q$ (all the details are given in SM). The fits of the PL time dependence are shown in figure \ref{fig:trplFEL}(b) and \ref{fig:trplFEL}(d). This way we extract a recovery time after the FEL pulse of $\tau_{cooling}=40\pm10\,$ps at $10\;$K of lattice temperature. The PL rise time is $\tau_{rise}= 50\pm10\;$ps.

At this point, the most important question to answer concerns the mechanism responsible for the PL quenching. As discussed in the literature for quantum well and quantum dot semiconductor systems, various physical mechanisms can lead to a PL quenching \cite{Rice2013,Klik2005,Zybell2011,Brotons-Gisbert2019,Cerne1996,Hughes1998}. 

To investigate whether we observe a resonant effect, we repeated the experiment at different FEL photon energies and we did not observe any significant dependence within our signal-to-noise ratio (see SM). We conclude that we observe a broadband non-resonant effect; thus, resonant effects such as phonon excitation \cite{Sekiguchi2021}, resonant excitation of shallow impurities \cite{Klik2005}, intra-excitonic transitions (e.g. 1s-to-2p) \cite{Rice2013,Poellmann2015}, and intersubband transitions \cite{Zybell2011} can be ruled out. 

We also rule out exciton ionization as origin of the PL quenching since the fluence dependence of the PL quenching shows a saturation behavior that is in contradiction with an exciton ionization process \cite{Brotons-Gisbert2019} and, moreover, we observe a strong PL quenching at THz fields of the order of $1\;$kV/cm which is much lower than the needed electric field for exciton ionization in InSe, since the exciton binding energy is of about $20\,$meV \cite{Cerne1995,Shubina2019,Felton2020}. 

Furthermore, lattice heating due to non-resonant absorption by the substrate could lead to a PL quenching but 1) we do not expect this fast dynamics, 2) the substrate is transparent and has a very high thermal conductivity (no heat transfer from the substrate to the flake and good thermal dissipation), and 3) the PL emission shows a slight blueshift after the THz pulse arrival while an increase of lattice temperature would induce a redshift of the PL emission \cite{Mudd2014, Christiansen2017}. 
 
\begin{figure*}
\centering
\includegraphics[scale=0.23]{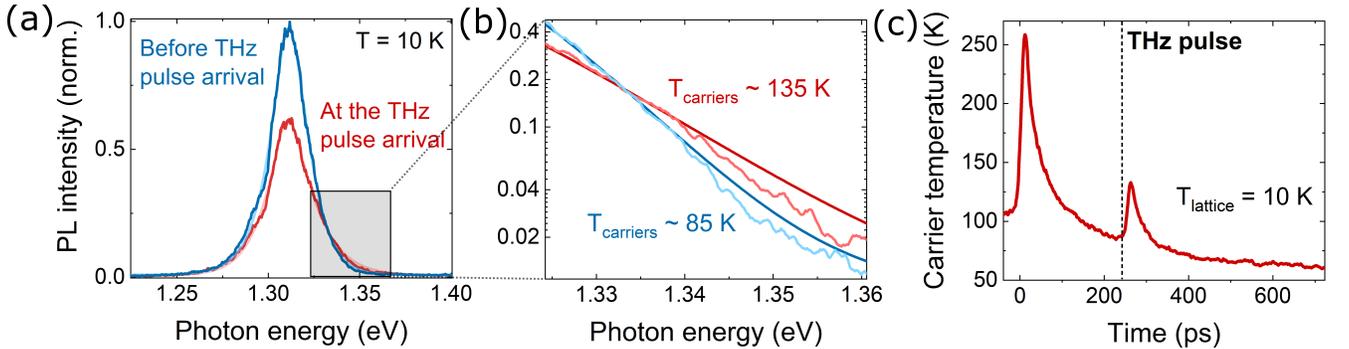}

\caption{(a) PL spectra before and during the arrival of the THz pulse. The PL emission is strongly quenched and a broadening of the peak on the high-energy side is observed. (b) Zoom in of the high-energy tail of the PL spectra. The emission gets broader at higher effective carrier temperatures. (c) Time dependence of the carrier temperature. The FEL pulse heats the carrier system and the radiative efficiency drops dramatically.}
\label{fig:trplSpec}
\end{figure*}

The most probable origin of the PL quenching is THz absorption by photo-excited free carriers. To verify this interpretation, we analyze the spectral shape of the PL emission as a function of time (shown in Figure \ref{fig:trplSpec}(a) and \ref{fig:trplSpec}(b)). We fit the spectrum at any time with a convolution of an Urbach tail $U(E)\propto \exp(\frac{E}{E_u})$ with the electronic density of state $D(E) \propto  \Theta(E-E_g)$ where $E_u$ is the Urbach energy, $E_g$ is the band-gap energy, and $\Theta$ is the Heaviside step function \cite{Katahara2014,Venanzi2020}. The fitting function was then weighted with a Fermi-Dirac distribution $f\propto (1+\exp(\frac{E-\mu}{k_B T}))^{-1}$. The resulting PL intensity reads as:
\begin{equation}
\label{eq:plFunction}
I_{PL}(\omega) \propto \omega^2 \Big( U(E)*D(E) \Big) f(\omega,T)
\end{equation} 

where $\omega$ is the photon frequency. The complete analytic expression is given in SM. We use equation (\ref{eq:plFunction}) to fit the spectrum at any delay time using two free parameters: the carrier temperature and the density of carriers that contribute to the PL emission. Figure \ref{fig:trplSpec}(a) shows the spectra with the fit at two delay times, i.e. just before and during the FEL pulse arrival. By comparing the two spectra it can be seen that the FEL pulse induces a quenching and a broadening on the high-energy side of the peak (see figure \ref{fig:trplSpec}(b)). This latter effect is direct evidence of hot-carrier emission \cite{Purschke2018}. Figure \ref{fig:trplSpec}(c) shows the time dependence of the carrier temperature as obtained from the fit. The carrier temperature shows a sharp increase after the visible pulse arrival due to the excitation of hot electron-hole pairs. A second sharp increase of the carrier temperature is observed at the THz pulse arrival. The increase of the carrier temperature is of about $50\,$K. From this value, we estimate a terahertz absorption of $A=0.12\%$ by the photo-excited carriers (see SM). 

THz absorption by free carriers explains also the saturation behavior of the PL quenching at high THz fluences shown in figure \ref{fig:trplFEL}(e). At high carrier temperatures, i.e. high THz fluences, the carrier cooling time becomes shorter and the electron heat capacity larger. The fast cooling of hot carriers via optical phonons leads to a limitation on the temperature that the carriers can reach and, therefore, a limitation on the maximum PL quenching. Similarly, at low temperature the electron heat capacity increases linearly with the temperature and this further reduces the heating of the carrier distribution at high THz fluences. A detailed discussion of these mechanisms is given in SM alongside with another possible mechanism that leads to a saturation behavior at high THz fluences. 

Now, we show how the heating of the photo-excited carriers by THz radiation leads to a PL quenching. To this end, we calculate the PL emission intensity with a microscopic theory assuming a quasi-stationary carrier distribution. Therefore, we calculate the luminescence in the free particle limit \cite{Kira1999}:
\begin{equation}
I(\Omega_{q_z}) \propto \im \left( d \sum_\mathbf{k} \frac{  f^e_\mathbf{k}(t) f^h_\mathbf{k}(t)}{E_g +\frac{\hbar^2 \mathbf{k}^2}{2 m}- \hbar\Omega_{q_z} - i \gamma} \right),\label{eq:lumi}
\end{equation}
with the dipole strength $d$, carrier momentum $\mathbf{k}$, band-gap energy $E_g$, reduced mass of the electron-hole pair $m$ and a phenomenological chosen broadening $\gamma$. Note that we assume the optical dipole element \textit{d} to be independent of the momentum, which is reasonable due to the negligible momenta of the photons \cite{Kira1999}. $f^e_\mathbf{k}(t)$ and $f^h_\mathbf{k}(t)$ are the time-dependent electron and hole distribution after the heating process, respectively.

To obtain the luminescence from equation \ref{eq:lumi}, we calculate the electron and hole distribution ($f^e_\mathbf{k}$ and $f^h_\mathbf{k}$) dynamics with the Boltzmann scattering equation in the low density limit:
\begin{widetext}
\begin{align}
\frac{d}{dt} f^{\lambda i}_\mathbf{q} = & \frac{2\pi}{\hbar} \sum_{\mathbf{k},j,\alpha,\pm} |g_{\mathbf{k-q}}^{\lambda\alpha} |^2 \left( \frac{1}{2} \pm \frac{1}{2} + n_\mathbf{k-q}^\alpha \right) \delta \left( \epsilon^i_\mathbf{q} - \epsilon^j_\mathbf{k} \pm \hbar \omega^\alpha_\mathbf{k-q} \right)  f^{\lambda j}_\mathbf{k} - 
\nonumber \\
& - \frac{2\pi}{\hbar}\sum_{\mathbf{k},\alpha,\pm} |g_{\mathbf{k-q}}^{\lambda\alpha} |^2 \left( \frac{1}{2} \pm \frac{1}{2} + n_\mathbf{k-q}^\alpha \right)\delta \left( \epsilon^i_\mathbf{q} - \epsilon^j_\mathbf{k} \mp \hbar \omega^\alpha_\mathbf{k-q} \right)   f^{\lambda i}_\mathbf{q}.\label{boltzmann}
\end{align}
\end{widetext}

The equation accounts for the thermalization of electrons to the lattice temperature, which is encoded in the phonon occupation $n^\alpha_\mathbf{q}$. The summation $\pm$ accounts for phonon emission $(+)$ and phonon absorption $(-)$ processes which are involved in the scattering. $n^\alpha_\mathbf{q}$ accounts for the phonon occupation of the mode $\alpha$ with momentum $\mathbf{q}$. The Dirac delta function imposes the energy conservation during the electron-phonon scattering event. $\epsilon^{\lambda i}_\mathbf{k} = \frac{\hbar^2 \mathbf{k}^2}{2 m^\lambda}$ is the carrier dispersion as a function of the carrier momentum $\mathbf{k}$, band $\lambda$, and the electronic subband $i$. The intersubband spacing is calculated in SM and is about $28\;$meV between the first and second subband. The effective masses in the conduction and the valence band are $m^c = 0.12 m_0$ and $m^v =  0.73 m_0$, respectively \cite{Mudd2016,Ferrer1997}. $g_\mathbf{q}^{\alpha,\lambda}$ is the electron-phonon coupling element. Further details are given in SM.

We solve the Boltzmann equation (Eq. \ref{boltzmann}) for lattice temperature $T_{lattice}=10\;$K and by setting an initial carrier temperature $T_{add}=50\,$K as obtained by the spectral fit of the PL emission just after the FEL pulse arrival. We note that we perform the calculation for different $T_{add}$ and show in SM that the results do not change qualitatively by varying this parameter. With this procedure, we are able to calculate the electron and hole cooling time as a function of carrier temperature and, therefore, the luminescence intensity as a function of time.

\begin{figure*}
\centering
\includegraphics[scale=0.30]{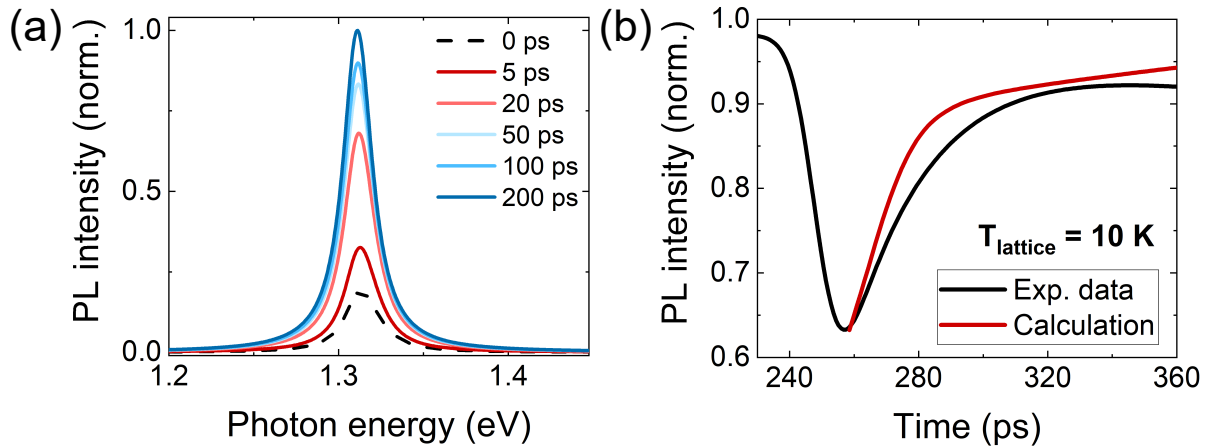}

\caption{(a) Calculated PL spectra as a function of delay time after the THz pulse arrival on the sample. After the initial quenching, the PL emission increases due to the relaxation of the heated electrons (holes) to the bottom (top) of the conduction (valence) band. (b) Comparison between experimental and calculated PL recovery after the THz pulse arrival at $10\;$K.}
\label{fig:calcMalte}
\end{figure*}

Figure \ref{fig:calcMalte}(a) shows the calculated PL spectra for selected times after the FEL pulse arrival. A PL recovery is clearly observed as consequence of the carrier cooling while the carrier density is kept constant. The origin of the quenching relates to the broadening of the electron and hole distributions when the carrier temperature increases. This can be understood by considering equation \ref{eq:lumi}: when the carrier temperature increases, the carrier distributions are broader in the momentum space and the sum over \textbf{k} in the numerator strongly reduces leading to a PL quenching.

We stress that the denominator in Eq. \ref{eq:lumi} contributes to shift the PL emission to higher energies. This observation is in agreement with the experimental data. However, the denominator does not have a significant influence on the PL quenching that is instead ruled by the numerator. The following is the intuitive interpretation of the observed PL quenching: the band-to-band bimolecular recombination depends on the probability to find both an electron and a hole at \textbf{k} and this probability is strongly reduced when the electron and hole distributions become broader in the momentum space. Further details are given in SM.

With this calculation, we can compare the time dependence of the PL recovery after the THz pulse arrival with the experimental data (see Figure \ref{fig:calcMalte}(b) for a lattice temperature of $10\;$K). We note that in figure \ref{fig:calcMalte}(b), we multiply the calculated curve by a multiplication factor to match the intensity of the PL quenching. In fact, we notice that from the theoretical calculation we expect a PL quenching of about $70\%$ (Figure \ref{fig:calcMalte}(a)) differently from the experimentally observed quenching of $35\%$. The reasons for this quantitative disagreement are 1) the limited time resolution of the setup that smear out the dip in the recorded PL spectra and make the experimental value of quenching smaller than the actual quenching; 2) in the calculation, we made the approximation of a quasi-stationary distribution so that we could define a carrier temperature at any time. This approximation can be critical especially during the arrival of the THz pulse and can lead to a wrong value of the calculated PL quenching. 

\begin{figure}
\centering
\includegraphics[scale=0.29]{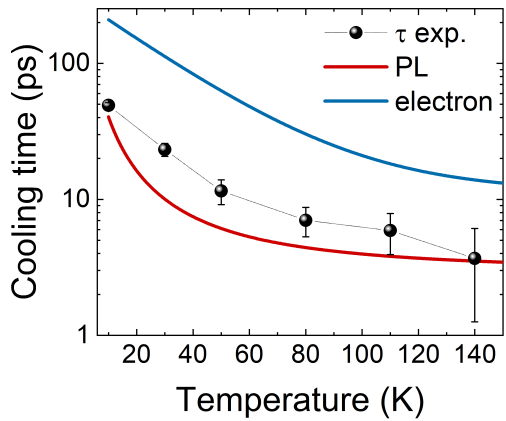}

\caption{Lattice temperature dependence of measured (black dots) and theoretically predicted (red line) PL recovery time. For comparison we also show the electron cooling time (blue line), which is quite different from the calculated PL recovery time as explained in the main text. The measured temperatures are $10\,$K, $30\,$K, $50\,$K, $80\,$K, $110\,$K, and $140\,$K.}
\label{fig:Tdep}
\end{figure}

To further corroborate our results and to get more information on the material properties, we investigate the dependence of the PL quenching on the lattice temperature. Figure \ref{fig:Tdep} shows the lattice temperature dependence of the PL recovery time (black dots). The error bars are calculated considering the time resolution of the setup and the error of the fitting procedure. The strength of the PL modulation decreases at high temperatures (data in SM) as consequence of higher electron heat capacity and shorter relaxation time. Figure \ref{fig:Tdep} shows also the PL cooling times calculated with the microscopic theory (red line). By comparing the experimental and calculated PL recovery time, we find a good agreement between calculation and data. This strongly suggests that our interpretation and our theory capture correctly the physics behind the modulation of the PL signal by the THz pulse.

Interestingly, the cooling time determined via a Boltzmann fitting of the electron occupation is one order of magnitude longer than the measured data (Fig. \ref{fig:Tdep}, blue line). This is connected to the missing momentum resolution in the simplification of the complex carrier relaxation under the influence of electron-phonon coupling to a single number (the cooling time). Differently from our full microscopic model, this method to obtain the relaxation time does not account for the fact that the PL emission arises mostly from occupied states with small momenta. In fact, electron-hole bimolecular luminescence is sensitive to the most densely populated regions, which are the low \textbf{q}-regions because they are quickly populated via optical phonon scattering.

As discussed, we have used Eq.\ref{boltzmann} to calculate, ab-initio, the temperature dependence of the cooling times finding a very good agreement with the experimental data. If we start from our calculation and leave the electron-phonon coupling constant $g$ as free parameter (not done in any plotted calculation), we find that the best fit is obtained for $g'=0.88g$, a slightly smaller value of the coupling constant than what expected from DFT calculations \cite{Shi2019,Wang2019first}.

The temperature dependence of the cooling time can mainly be understood from the scattering of electrons by optical phonons. At low temperatures, optical phonons only have a minor impact on the relaxation because the optical phonon energy ($\hbar \omega_{ZO_1}=12.4\,$meV \cite{Shi2019}) is too large to contribute effectively to the cooling of carriers with low kinetic energy \cite{Murdin1997}. Therefore, the cooling is mainly determined via acoustic phonon scattering at low temperatures. Increasing the lattice temperature increases the kinetic energies of electrons resulting in a larger impact of optical phonon scattering to the cooling dynamics. This leads to a speed-up of the cooling at high lattice temperatures.

Finally, we note that the strength of the PL quenching depends mostly on 1) THz absorption, 2) electron heat capacity, and 3) carrier relaxation time. In other words, it depends on how well the heating of the electron distribution via THz absorption works. In this regard, the low electron mass plays a central role since the THz absorption and the heat capacity depends on the electron mass ($A\sim \frac{n}{m}$ and $c \sim m$ in a two-dimensional electron gas, respectively). High absorption and low heat capacity lead to a high heating of the carrier distribution and, consequently, to a strong PL quenching. The carrier relaxation time plays also an important role and long time corresponds to high quenching. To conclude, the PL quenching is expected in other vdW semiconductors and the effect will be particularly strong for materials that have low carrier masses and long relaxation time, which is the case for InSe.

In summary, we investigated the effects of THz radiation on the PL emission of few-layer InSe. We observe a transient quenching of the PL emission induced by heating of the photo-excited electron-hole system. The PL quenching reaches values up to $52\%$. The heating and cooling of the electron-hole system are modeled by Boltzmann equation and the PL emission is calculated microscopically. With the microscopic theory, we are able to reproduce the experimental data corroborating the physical interpretation that THz free carrier absorption leads to a broader carrier distribution in the momentum space and, therefore, to a PL quenching. The increase of cooling time at low temperatures is attributed to the less efficient cooling via emission of optical phonons because of their high energy with respect to the thermal energy of carriers. By comparing the experimental and theoretical results, we provide an experimental estimation of the electron-phonon coupling constants of $0.88$ times of the expected values from DFT calculations.

All in all, we demonstrate the possibility to use terahertz radiation for the dynamic control of the photoluminescence emission in a van der Waals semiconductor. This work demonstrates a promising route for the realization of opto-electronic technology, such as terahertz detectors and optical modulators.

\section*{Supplementary Material}

Supplementary material with further experiments and calculations is provided.

\begin{acknowledgments}
The authors thank A. Chernikov and M. Ortolani for helpful and friendly discussions. M.S., M.K and A.K. gratefully acknowledge funding by the Deutsche Forschungsgemeinschaft (DFG) - project number 182087777 - SFB 951. 
\end{acknowledgments}

\section*{Data Availability Statement}

The data that support the findings of this study are openly available in Rodare, at https://rodare.hzdr.de/. Raw data were generated at Free-electron laser FELBE at HZDR. Derived data supporting the findings of this study are available from the corresponding author upon reasonable request.


\bibliography{mybib}
\end{document}